\documentclass[aps,prl,twocolumn,preprintnumbers,
               showpacs,nofootinbib]{revtex4}
\newcommand{\PRE}[1]{}       

\usepackage{bm}
\usepackage{epsfig}

\newcommand{\postscript}[2]{\setlength{\epsfxsize}{#2\hsize}
   \centerline{\epsfbox{#1}}}

\newcommand{\OmegaDM}{\Omega_{\text{DM}}}

\newcommand{\kev}{\text{keV}}
\newcommand{\mev}{\text{MeV}}
\newcommand{\gev}{\text{GeV}}
\newcommand{\tev}{\text{TeV}}

\newcommand{\cm}{\text{cm}}

\newcommand{\s}{\text{s}}

\newcommand{\sr}{\text{sr}}
\newcommand{\etal}{{\em et al.}}
\newcommand{\eg}{{\em e.g.}}

\newcommand{\eqref}[1]{Eq.~(\ref{#1})}

\newcommand{\WIMP}{\text{WIMP}}
\newcommand{\SWIMP}{\text{SWIMP}}
\newcommand{\mWIMP}{m_{\WIMP}}
\newcommand{\mSWIMP}{m_{\SWIMP}}

\begin{document}

\preprint{UCI-TR-2003-6}

\title{
\PRE{\vspace*{1.5in}}
Superweakly Interacting Massive Particles
\PRE{\vspace*{0.3in}}
}

\author{Jonathan L.~Feng}
\affiliation{Department of Physics and Astronomy,
University of California, Irvine, CA 92697, USA
\PRE{\vspace*{.1in}}
}
\author{Arvind Rajaraman}
\affiliation{Department of Physics and Astronomy,
University of California, Irvine, CA 92697, USA
\PRE{\vspace*{.1in}}
}

\author{Fumihiro Takayama%
\PRE{\vspace*{.2in}}
}
\affiliation{Department of Physics and Astronomy,
University of California, Irvine, CA 92697, USA
\PRE{\vspace*{.1in}}
}

\date{February 2003}

\begin{abstract}
\PRE{\vspace*{.1in}} We investigate a new class of dark matter:
superweakly-interacting massive particles (superWIMPs).  As with
conventional WIMPs, superWIMPs appear in well-motivated particle
theories with naturally the correct relic density.  In contrast to
WIMPs, however, superWIMPs are impossible to detect in all
conventional dark matter searches.  We consider the concrete examples
of gravitino and graviton cold dark matter in models with
supersymmetry and universal extra dimensions, respectively, and show
that superWIMP dark matter satisfies stringent constraints from Big
Bang nucleosynthesis and the cosmic microwave background.
\end{abstract}


\maketitle

There is ample evidence that luminous matter makes up only a small
fraction of all matter in the universe. Results from the Wilkinson
Microwave Anisotropy Probe, combined with other data, constrain the
non-baryonic dark matter density to $\OmegaDM = 0.23 \pm
0.04$~\cite{Spergel:2003cb}, far in excess of the luminous matter
density. We therefore live in interesting times: while the amount of
dark matter is becoming precisely known, its identity remains a
mystery.

WIMPs, weakly-interacting massive particles with weak-scale masses,
are particularly attractive dark matter candidates.  WIMPs have
several virtues.  First, their appearance in particle physics theories
is independently motivated by the problem of electroweak symmetry
breaking.  Second, given standard cosmological assumptions, their
thermal relic abundance is naturally that required for dark matter.
Third, the requirement that WIMPs annihilate efficiently enough to
give the desired relic density generically implies that WIMP-matter
interactions are strong enough for dark matter to be discovered in
current or near future experiments.

Here we consider a new class of non-baryonic cold dark matter:
superweakly-interacting massive particles (superWIMPs or SWIMPs).  As
with WIMPs, superWIMPs appear in well-motivated theoretical
frameworks, such as supersymmetry and extra dimensions, and their
(non-thermal) relic density is also naturally in the desired range. In
contrast to conventional WIMPs, however, they interact superweakly and
so evade all direct and indirect dark matter detection experiments
proposed to date.

For concreteness, we consider two specific superWIMPs: gravitinos in
supersymmetric theories, and Kaluza-Klein (KK) gravitons in theories
with extra dimensions.  Gravitino and graviton superWIMPs share many
features, and we investigate them in parallel.

For gravitino superWIMPs, we consider supergravity, where the
gravitino $\tilde{G}$ and all standard model (SM) superpartners have
weak-scale masses.  Assuming $R$-parity conservation, the lightest
supersymmetric particle (LSP) is stable.  In supergravity, the LSP is
usually assumed to be a SM superpartner.  Neutralino LSPs are
excellent WIMP candidates, giving the desired thermal relic density
for masses of $50~\gev$ to $2~\tev$, depending on Higgsino content. In
contrast, here we assume a $\tilde{G}$ LSP. The gravitinos considered
here couple gravitationally and form cold dark matter, in contrast to
the case in low-scale supersymmetry breaking models where light
gravitinos couple more strongly and form warm dark matter.

We consider also the possibility of graviton dark matter in universal
extra dimensions (UED), in which gravity and all SM fields
propagate~\cite{Appelquist:2000nn}.  We focus on $D=5$ spacetime
dimensions with coordinates $x^M = (x^{\mu},y)$.  The fifth dimension
is compactified on the orbifold $S^1/Z_2$, where $S^1$ is a circle of
radius $R$, and $Z_2$ corresponds to $y \to -y$.  Unwanted massless
fields are removed by requiring suitable transformations under $y \to
-y$.  For example, the 5D gauge field $V_M(x,y)$ transforms as
$V_\mu(x,y) \to V_\mu(x,y)$ and $V_5(x,y) \to -V_5(x,y)$ under $y \to
-y$, which preserves $V_\mu^0(x)$ and removes $V_5^0(x)$. Similar
choices remove half of the fermionic degrees of freedom, producing
chiral 4D fermions, and preserve the 4D graviton $h_{\mu\nu}^0(x)$
while removing $h_{\mu 5}^0(x)$ and $h_{5 \nu}^0(x)$.  The
gravi-scalar $h_{55}^0(x)$ remains; we assume that some other physics
stabilizes this mode and generates a mass for it.

The orbifold compactification breaks KK number conservation, but
preserves KK-parity.  KK particles must therefore be produced in
pairs, and current bounds require only $R^{-1} \agt
200~\gev$~\cite{Appelquist:2000nn,Appelquist:2002wb}.  KK-parity
conservation also makes the lightest KK particle (LKP) stable and a
dark matter candidate.  For $R^{-1} \sim \tev$, weakly-interacting KK
particles have thermal relic densities consistent with
observations~\cite{Servant:2002aq}. In particular, $B^1$, the first KK
partner of the U(1) gauge boson, has been shown to be a viable WIMP
dark matter candidate, with promising prospects for direct
detection~\cite{Cheng:2002ej,Servant:2002hb} and also indirect
detection in anti-matter searches~\cite{Cheng:2002ej}, neutrino
telescopes~\cite{Cheng:2002ej,Hooper:2002gs,Bertone:2002ms}, and gamma
ray detectors~\cite{Cheng:2002ej,Bertone:2002ms}.

As in the case of supersymmetry, however, the lightest partner need
not be a SM partner.  In UED, the LKP could be $G^1$, the first KK
partner of the graviton.  $G^1$ is, in fact, perhaps the most natural
LKP candidate --- radiative contributions to KK masses, typically
positive~\cite{Cheng:2002iz}, are negligible for $G^1$. $G^1$
couplings are also gravitational, and so highly suppressed.

Gravitinos and gravitons therefore naturally emerge as superWIMP
candidates: stable massive particles with superweak interactions.
Their weak gravitational interactions imply that they play no role in
the thermal history of the early universe. (We assume inflation
followed by reheating to a temperature low enough to avoid
regenerating large numbers of superWIMPs.)  Thus, if the next lightest
supersymmetric particle (NLSP) or next lightest KK particle (NLKP) is
weakly-interacting, it freezes out with a relic density of the desired
magnitude.  Much later, however, these WIMPs then decay to superWIMPs;
as the WIMP and superWIMP masses are similar, the superWIMP then
inherits the desired relic density.

Unlike WIMPs, however, superWIMPs are impossible to discover directly,
and their annihilation rate is so suppressed that they also escape all
indirect detection experiments.  At the same time, unlike superheavy
dark matter candidates with only gravitational
interactions~\cite{Chung:2001cb}, superWIMPs inherit the desired relic
density from a thermal abundance and arise from accessible electroweak
physics. At colliders, WIMP decays to superWIMPs will occur long after
the WIMP leaves the detector.  If the NLSP or NLKP is neutral, its
metastability will have no observable consequences.  The discovery of
a seemingly stable but charged NLSP or NLKP may, however, provide a
strong hint for superWIMP dark matter.

We now investigate constraints on and alternative signals of superWIMP
dark matter scenarios.  The observable consequences of superWIMPs must
rely on the decays of WIMPs to superWIMPs on cosmological time
scales~\cite{Ellis:1984er}. The NLSP or NLKP may be any SM partner.
In supergravity, the lightest SM superpartner is often the Bino
$\tilde{B}$, the superpartner of the hypercharge gauge boson.  In the
minimal UED scenario~\cite{Appelquist:2000nn,Cheng:2002iz}, the
lightest SM KK mode is often $B^1$.  Motivated by these results, we
now consider specific scenarios in which decays to superWIMPs are
typically accompanied by photons, and we consider the impact of
electromagnetic cascades.

In the supersymmetric photon superWIMP scenario, NLSP decay is
governed by the coupling $- \frac{i}{8 M_*} \bar{\tilde{G}}_{\mu}
\left[ \gamma^{\nu}, \gamma^{\rho} \right] \gamma^{\mu} \tilde{B} \,
F_{\nu\rho}$, where $F$ is the U(1) field strength, and $M_* = (8\pi
G_N)^{-1/2} \simeq 2.4 \times 10^{18}~\gev$ is the reduced Planck
scale.  The NLSP decay width is
\begin{eqnarray}
\!\!\! \Gamma(\tilde{B} \to \tilde{G} \gamma) \!\!
&=&  \!\! \frac{\cos^2\theta_W}{48\pi M_*^2}
\frac{m_{\tilde{B}}^5}{m_{\tilde{G}}^2} \!
\left[1 \! - \! \frac{m_{\tilde{G}}^2}{m_{\tilde{B}}^2} \right]^3 \!
\left[1 \! + \! 3 \frac{m_{\tilde{G}}^2}{m_{\tilde{B}}^2} \right] .
\label{Binolifetime}
\end{eqnarray}
In models with low-scale supersymmetry breaking, $m_{\tilde{G}} \ll
m_{\tilde{B}}$, and the gravitino couples dominantly through its $\pm
\frac{1}{2}$ spin components.  In the high-scale supersymmetry
breaking scenarios considered here, however, the couplings of the $\pm
\frac{3}{2}$ spin polarizations are of the same order and must be kept
in deriving \eqref{Binolifetime}.

The properties of gravitons in UED scenarios may be determined
straightforwardly; details will be presented elsewhere~\cite{inprep}.
Graviton superWIMPs couple to $B^1$ through $\frac{\sqrt{2}}{M_*}
G^1_{\mu\nu} ( -F^{0\, \mu\rho} F^{1 \nu}{}_\rho + \frac{1}{4}
\eta^{\mu\nu} F^0_{\rho\sigma} F^{1 \rho\sigma} )$, where $F^n_{\mu
\nu} \equiv\partial_{\mu} B^n_{\nu} - \partial_{\nu} B^n_{\mu}$.  The
$G^1 B^1 B^0$ vertex is identical to the $G^0 B^0 B^0$ vertex, and the
longitudinal component of the massive $B^1$ plays no role.  The NLKP
decay width is
\begin{eqnarray}
 \Gamma(B^1 \to G^1 \gamma)
&=&  \frac{\cos^2\theta_W}{72\pi M_*^2}
\frac{m_{B^1}^7}{m_{G^1}^4}
\left[1 - \frac{m_{G^1}^2}{m_{B^1}^2} \right]^3 \nonumber \\
&& \times \left[1 + 3 \frac{m_{G^1}^2}{m_{B^1}^2}
 + 6 \frac{m_{G^1}^4}{m_{B^1}^4} \right] .
\label{B1lifetime}
\end{eqnarray}

\begin{figure}[tbp]
\postscript{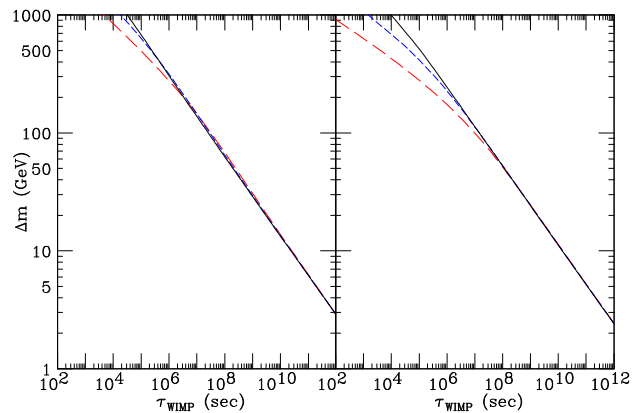}{0.95}
\caption{Lifetimes for $\tilde{B} \to \tilde{G} \gamma$ (left) and
$B^1 \to G^1 \gamma$ (right) for $\Delta m \equiv \mWIMP - \mSWIMP$
and $\mSWIMP = 0.1~\tev$ (long dashed), $0.3~\tev$ (short dashed), and
$1~\tev$ (solid).
\label{fig:lifetime} }
\end{figure}

The $\tilde{B}$ and $B^1$ lifetimes are given in
Fig.~\ref{fig:lifetime}.  In both cases, in the limit $\Delta m \equiv
\mWIMP - \mSWIMP \ll \mSWIMP$, the WIMP lifetime is proportional to
$(\Delta m)^{-3}$ and is independent of the overall mass scale.

\begin{figure}[tbp]
\postscript{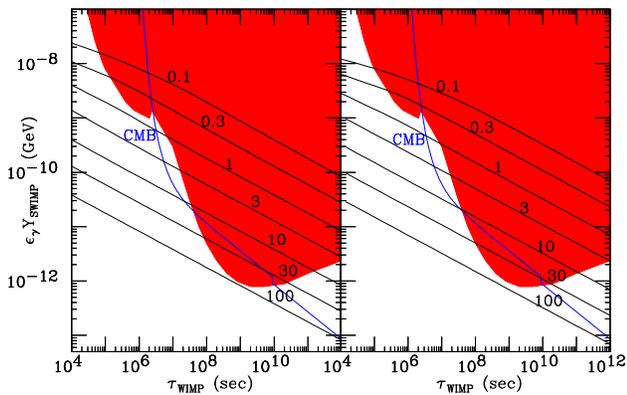}{0.95}
\caption{The photon energy release $\varepsilon_\gamma Y_{\SWIMP}$ for
various $\mSWIMP$ in TeV in the gravitino (left) and graviton (right)
superWIMP scenarios. We fix $\Omega_{\text{SWIMP}} = 0.23$;
$\varepsilon_\gamma Y_{\SWIMP}$ scales linearly with
$\Omega_{\text{SWIMP}}$. BBN constraints exclude the shaded
regions~\cite{Cyburt:2002uv}; consistency of the CMB with a black-body
spectrum excludes regions above the CMB contours.
\label{fig:bbn} }
\end{figure}

As we will see, the most relevant bounds constrain the total energy
released in photons in WIMP decay, or more precisely,
$\varepsilon_\gamma Y_\gamma$, where $\varepsilon_\gamma$ is the
energy of the photons when created and $Y_\gamma =
n_{\gamma}/n_{\gamma}^{\text{BG}}$ is the number density of photons
from WIMP decay normalized to the number density of background photons
$n_{\gamma}^{\text{BG}} = 2 \zeta(3) T^3/\pi^2$, where $T$ is the
temperature during WIMP decay.  In the superWIMP scenario, WIMPs decay
essentially at rest, and so $\varepsilon_\gamma = (\mWIMP^2 -
\mSWIMP^2)/(2\mWIMP)$.  Since a superWIMP is produced in association
with each photon, $Y_\gamma = Y_{\SWIMP}$, and the photon abundance is
given by
\begin{equation}
Y_{\SWIMP} \simeq 3.0 \times 10^{-12}
\left[\frac{\tev}{m_{\SWIMP}}\right]
\left[\frac{\Omega_{\SWIMP}}{0.23}\right] .
\label{predictedabundance}
\end{equation}
Predicted values for $\varepsilon_\gamma Y_{\SWIMP}$ are shown in
Fig.~\ref{fig:bbn}.

As evident in Fig.~\ref{fig:lifetime}, WIMP decays occur long after
Big Bang nucleosynthesis (BBN), and so may, in principle, destroy the
successful BBN predictions for light element abundances.  The energy
of photons produced in late decays is rapidly redistributed through
scattering off background photons, $\gamma \gamma_{\text{BG}} \to e^+
e^-$, and inverse Compton
scattering~\cite{Kawasaki:1994sc,Cyburt:2002uv}.  As a result, the
constraints of BBN are, to an excellent approximation, independent of
the initial energy distribution of injected photons, and constrain
only the total energy release.

Detailed analysis~\cite{Cyburt:2002uv}, demanding consistent
predictions for deuterium, $^3$He, $^4$He, $^6$Li, and $^7$Li,
excludes the region of parameter space shown in Fig.~\ref{fig:bbn}.
The BBN constraint is weak for early decays: at early times, the
universe is hot and the initial photon spectrum is rapidly softened,
leaving few high energy photons above threshold to modify the light
element abundances.  We find that BBN excludes some of the relevant
parameter region, but not all.  In particular, for relatively
short-lived WIMPs with $\tau \alt 10^7~\s$ and weak-scale masses, the
requirement that superWIMPs form all of the dark matter is consistent
with the constraints from BBN.

The cosmic microwave background (CMB) also imposes
constraints~\cite{Hu:gc,Fixsen:1996nj}.  The injection of energy in
the form of photons may distort the CMB from the observed black-body
spectrum.  Before redshifts of $z \sim 10^7$, elastic Compton
scattering, bremsstrahlung $e X \to e X \gamma$ (with $X$ an ion), and
double Compton scattering $e^- \gamma \to e^- \gamma \gamma$
effectively thermalize injected energy.  After $z \sim 10^7$, however,
the photon number-changing interactions become ineffective, and the
photon spectrum relaxes only to a Bose-Einstein distribution with
chemical potential $\mu$.  After $z \sim 10^5$, even Compton
scattering becomes ineffective, and deviations from the black-body
spectrum may be parametrized by the Sunyaev-Zeldovich $y$ parameter.

As with the BBN constraints, bounds from the CMB are largely
independent of the injected energy spectrum, depending primarily on
the total energy release.  The bounds on $\varepsilon_\gamma
Y_{\SWIMP}$ scale linearly with the bounds on $\mu$ and $y$.  We
update the analysis of Ref.~\cite{Hu:gc} to include the latest results
$|\mu| < 9 \times 10^{-5}$ and $|y| < 1.2 \times
10^{-5}$~\cite{Hagiwara:fs}, with baryon density $\Omega_B h^2 \simeq
0.022$~\cite{Spergel:2003cb}, where $h \simeq 0.71$ is the normalized
Hubble expansion rate.  These bounds exclude energy releases above the
CMB contours in Fig.~\ref{fig:bbn}.  Remarkably, the CMB constraints
are now so precise that they supersede BBN constraints for decay times
$\tau \sim 10^7~\s$ and $\tau \agt 10^{10}~\s$.  Nevertheless, regions
of superWIMP parameter space, including regions with weak-scale masses
and mass splittings, remain viable.

Finally, for highly degenerate WIMP-SWIMP pairs, WIMPs decay very late
to soft photons.  The photon spectrum is not thermalized and may
produce observable peaks in the diffuse photon spectrum.  The present
differential flux of photons from WIMP decay is
\begin{equation}
\frac{d\Phi}{ dE_{\gamma}}
=\frac{c}{4\pi} \int_0^{t_0} \frac{dt}{\tau_{\WIMP}}
\frac{N_{\WIMP}(t)}{V_0}
\delta \left( E_{\gamma} - \frac{\varepsilon_\gamma}{1+z} \right) ,
\end{equation}
where $t_0 \simeq 13.7~\text{Gyr}$ is the age of the
universe~\cite{Spergel:2003cb}, $N_{\WIMP}(t) = N_{\WIMP}^{\text{in}}
e^{-t/\tau_{\WIMP}}$, where $N_{\WIMP}^{\text{in}}$ is the number of
WIMPs at freeze-out, and $V_0$ is the present volume of the
universe. The diffuse photon flux is a sensitive probe only when WIMPs
decay in the matter-dominated era.  We may then take $1+z =
(t_0/t)^{2/3}$, and
\begin{equation}
\frac{d\Phi}{ dE_{\gamma}} \simeq
\frac{3c}{8\pi} \frac{N_{\WIMP}^{\text{in}}}{V_0 \varepsilon_\gamma}
\left[\frac{t_0}{\tau_{\WIMP}}\right]^{\frac{2}{3}}
F(a) \, \Theta(\varepsilon_\gamma - E_{\gamma}) \ ,
\label{phi}
\end{equation}
where $F(a) = a^{1/2}e^{-a^{3/2}}$, $a = (E_{\gamma} /
\varepsilon_{\gamma}) (t_0/\tau_{\WIMP})^{2/3}$.  $F(a)$ is maximal at
$a = 3^{-2/3}$, and so, for $\tau_{\WIMP} < t_0$, the differential
photon flux reaches its maximal value at
\begin{equation}
E_{\gamma}^{\text{max}} = \varepsilon_{\gamma}
\left[ \frac{\tau_{\WIMP}}{3t_0}\right]^{\frac{2}{3}}
= 680~\kev \left[ \frac{\gev}{\Delta m} \right]
\end{equation}
for gravitinos, and an energy 1.4 times smaller for gravitons.  The
$\Delta m$ dependence follows from the redshifting of photons created
with energy $\Delta m$ by $1+z \propto \tau_{\WIMP}^{-2/3} \propto
(\Delta m)^2$.  For high degeneracies,
\begin{equation}
\frac{N_{\WIMP}^{\text{in}}}{V_0} = 1.2 \times 10^{-9}~\cm^{-3}
\left[ \frac{\tev}{m_{\SWIMP}} \right]
\left[ \frac{\Omega_{\SWIMP}}{0.23} \right] ,
\end{equation}
and so the maximal flux is
\begin{eqnarray}
\frac{d\Phi}{dE_{\gamma}} (E_\gamma^{\text{max}})
&=& 1.5~\cm^{-2}~\s^{-1}~\sr^{-1}~\mev^{-1} \nonumber \\
&&\times \left[ \frac{\tev}{m_{\SWIMP}} \right]
\left[ \frac{\Delta m}{\gev} \right]
\left[ \frac{\Omega_{\SWIMP}}{0.23} \right]
\label{phimax}
\end{eqnarray}
for gravitinos, and a factor of 1.4 larger for gravitons.

\begin{figure}[tbp]
\postscript{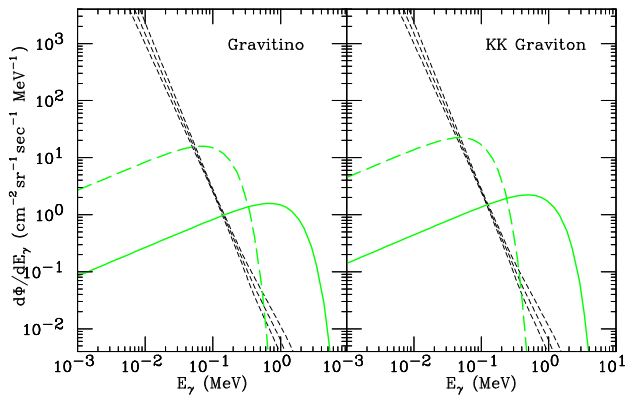}{0.95}
\caption{Diffuse photon fluxes (solid) for $\mSWIMP = 1~\tev$,
$\Omega_{\SWIMP} = 0.23$, and $\Delta m = 1~\gev$ (solid) and
$10~\gev$ (long dashed), and upper bounds from observations (short
dashed).
\label{fig:diffuseflux} }
\end{figure}

Representative photon energy spectra are shown in
Fig.~\ref{fig:diffuseflux}.  Also shown are the measured diffuse
fluxes from the observatories HEAO, OSSE, and COMPTEL~\cite{HEAO},
determined from observed fluxes by subtracting known point sources.
The observed spectra fall rapidly with energy and so severely
constrain the relatively hard photon spectra predicted by $\Delta m
\alt 10~\gev$.  Note that we have not included photon interactions
which soften the photon spectrum for $\Delta m \agt$ few
GeV~\cite{Asaka:1998ju}; such effects can only enlarge the allowed
parameter space discussed below.

\begin{figure}[tbp]
\postscript{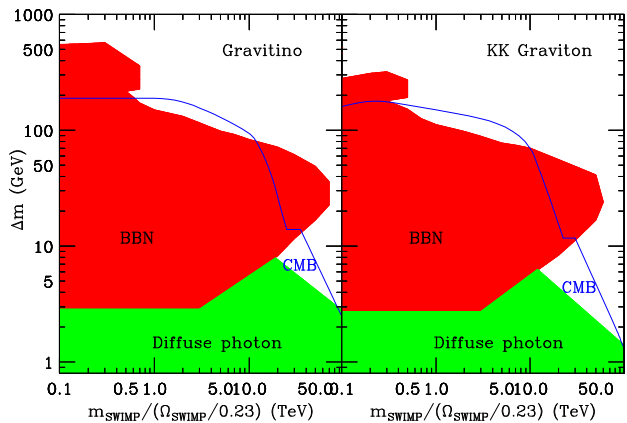}{0.95}
\caption{Regions of the $(\mSWIMP, \Delta m)$ plane excluded by BBN,
CMB, and diffuse photon constraints.  The shaded regions and the
regions below the CMB contours are excluded.
\label{fig:photonsummary} }
\end{figure}

In Fig.~\ref{fig:photonsummary}, we compile the constraints discussed
above and show the allowed regions of the $(\mSWIMP, \Delta m)$ plane
for the photon gravitino and graviton superWIMP dark matter scenarios.
The BBN and CMB constraints are as discussed above.  An additional
region is excluded by the requirement that the diffuse photon flux
never exceed the observed flux by $2\sigma$ for any energy.  Although
these data exclude some of the parameter space, the most
well-motivated region with $\mSWIMP, \Delta m \sim 100~\gev$ to
$1~\tev$ remains an outstanding possibility.

In conclusion, we find that superWIMPs provide a qualitatively novel
possibility for particle dark matter.  Such particles appear in the
form of gravitinos and gravitons in theories with supersymmetry and
extra dimensions, and they naturally inherit the desired relic density
from late-decaying weakly-interacting NLSPs or NLKPs.

SuperWIMPs satisfy existing constraints from BBN and CMB and evade all
conventional dark matter experiments. On the surface, we have
apparently committed Pauli's ``ultimate sin'' by proposing a solution
to the dark matter problem that has no observable consequences.
However, improvements in the measurements discussed above may uncover
anomalies.  For example, a detailed study of the $15~\kev$ to
$10~\mev$ diffuse photon flux by INTEGRAL is currently
underway~\cite{INTEGRAL}.  A pronounced bump in this spectrum could
provide a striking signal of superWIMP dark matter.  Finally, we note
that neutrino, charged lepton, weak gauge boson, and Higgs boson NLSPs
and NLKPs are all viable from the point of view of preserving the
naturalness of the desired relic density.  Some of these scenarios may
be severely constrained by bounds on hadronic showers, but the
remaining scenarios will have qualitatively different, and possibly
very interesting, observational consequences.

We thank J.~Cohn, T.~Han, T.~Moroi, and M.~White for enlightening
conversations.


\end{document}